\documentclass[a4paper,11pt]{article}
\pdfoutput=1 

\usepackage{jcappub} 

\usepackage[T1]{fontenc} 
\usepackage[numbers]{natbib}
\usepackage{graphicx,epsfig}

\title{\boldmath Constraining the range of Yukawa gravity interaction from S2
star orbits II: Bounds on graviton mass}

\author[a,b,c,d,e,1]{A. F. Zakharov,\note{Corresponding author.}}
\author[f]{P. Jovanovi\'{c},}
\author[g]{D. Borka}
\author[g]{and V. Borka Jovanovi\'{c}}

\affiliation[a]{National Astronomical Observatories of Chinese
Academy of Sciences,\\ Datun Road 20A, Beijing, 100012 China}
\affiliation[b]{Institute of Theoretical and Experimental Physics,
117259 Moscow, Russia}
\affiliation[c]{National Research Nuclear
University  MEPhI \\(Moscow Engineering Physics Institute), 115409,
Moscow,
 Russia}
\affiliation[d]{Bogoliubov Laboratory for Theoretical Physics, JINR,
141980 Dubna, Russia}
\affiliation[e]{North Carolina Central
University, Durham, NC 27707,
 USA}
\affiliation[f]{Astronomical Observatory, Volgina 7, 11060 Belgrade,
Serbia} \affiliation[g]{Atomic Physics Laboratory (040), Vin\v{c}a
Institute of Nuclear Sciences, \\University of Belgrade, P.O. Box
522, 11001 Belgrade, Serbia}
\emailAdd{zakharov@itep.ru} \emailAdd{pjovanovic@aob.rs}
\emailAdd{dusborka@vin.bg.ac.rs} \emailAdd{vborka@vin.bg.ac.rs}

\abstract{Recently LIGO collaboration discovered gravitational waves
\cite{Abbott_16} predicted 100 years ago by A. Einstein. Moreover,
in the key paper reporting about the discovery, the joint LIGO \&
VIRGO team presented an upper limit on graviton mass such as $m_g <
1.2 \times 10^{-22} eV$ \cite{Abbott_16} (see also more details in
another LIGO paper \cite{ligo16} dedicated to a data analysis to
obtain such a small constraint on a graviton mass). Since the
graviton mass limit is so small the authors concluded that their
observational data do not show violations of classical general
relativity.
  We consider another opportunity to evaluate a graviton mass from
 phenomenological consequences of massive
gravity and show that an analysis of bright star trajectories could
bound graviton mass with a  comparable accuracy with accuracies
reached with  gravitational wave interferometers and expected with
forthcoming pulsar timing observations for gravitational wave
detection. It gives an opportunity to treat observations of bright
stars near the Galactic Center as a wonderful tool not only for an
evaluation specific parameters of the black hole but also to obtain
constraints on the fundamental gravity law such as a modifications
of Newton gravity law in a weak field approximation. In particular,
we obtain bounds on a graviton mass based on a potential
reconstruction at the Galactic Center.}


\begin{document}
\maketitle
\flushbottom

\section{Introduction}

Recently long-term efforts of theorists and experimentalists have
been led to a remarkable discovery of gravitational waves
\cite{Abbott_16}. The result gives also  a confirmation of black
hole existence in binary systems. The discovery is also an
additional brilliant confirmation of general relativity. However,
alternative theories of gravity, including massive graviton theories
introduced by Fierz and Pauli \cite{FP} are a subject of intensive
studies and current and future experiments and observations can give
constraints on graviton mass. As it was noted by C. Will
\cite{Will_98,Will_14}, gravitational wave observations could
constrain  a graviton mass since in the case of massive graviton a
gravitational wave signal is different from a signal of general
relativity and analyzing differences for these two curves calculated
within these two models (massive and massless ones) one could obtain
a graviton mass constraint. Moreover, one could detect a time delay
of gravitational wave signal in respect of electromagnetic and
neutrino signals in the case if source of gravitational radiation
signal is known.

There are many alternative theories of gravity which have been
proposed in last years in spite of the evident success of the
conventional general relativity (GR) since after 100 years of its
development we do not have direct necessities to change GR with its
alternative. However, a slow progress in understanding of dark
matter (DM) and dark energy (DE) problems stimulates a growth of
interest to alternative theories of gravity. Perhaps, there are
deviation from Newtonian gravity at the Solar system scales
\cite{Anderson_98,Anderson_08} in spite of a model of the thermal
origin of the anomaly \cite{Turyshev_12}.

A version of a Lorentz invariant massive gravity has been introduced
by Fierz and Pauli \cite{FP}, however, later people found a number
of problems with such theories such as existence of ghosts, vDVZ
discontinuity \cite{Zakharov_70,van_Dam_70,Iwasaki_70} and some
other technical problems \cite{Rubakov_08}. Some of these problems
can be overcame
\cite{Rubakov_08,Vainshtein_72,Visser_98,Dvali_00,Kogan_01,Deffayet_02,Finn_02,Damour_03,Maggiorre_08,Hinterbichler_12}.

Nowadays some of alternative theories do not have the Newtonian
limit in a weak gravitational field approximation. Yukawa-like
corrections have been obtained, as a general feature, in the
framework of $f(R)$ theories of gravity \cite{capo09a}. A set of
gravity theories (including so-called massive graviton theories)
have a Yukawa limit for a weak gravitational field. We will discuss
observational consequences of massive gravity where one can expect a
Yukawa type of weak gravitational field limit instead of Newtonian
one.

Different experimental and observational ways to bound a graviton
mass have been suggested
\cite{Nieto_74,Freund_69,Logunov_84,Gershtein_06,Gershtein_08,Nieto_10}.
In particular, it was shown that speed of gravity practically
coincides with speed of light
\cite{Kopeikin_01,Kopeikin_03,Kopeikin_04,Kopeikin_05,Kopeikin_06},
so one could get constraints of a graviton mass. In paper
\cite{Talmadge_88} constraints on the range $\lambda$ of Yukawa type
force in Solar system has been considered and one can derive the
bound $\lambda > 2.8 \times 10^{12}$~km \cite{Will_14} from these
constraints assuming a natural modification of the Newtonian
potential \cite{Visser_98,Will_14}:
\begin{equation}
V \left( r \right) = -\dfrac{GM}{(1+\delta)r}\left[ {1 + \delta e^{-
\left(\dfrac{r}\lambda \right)}} \right], \label{eq_yukawa}
\end{equation}
where $\delta$ is a universal constant. In our previous paper
\cite{Dusko_JCAP_13} we found constraints on parameters of Yukawa
gravity.

Will considered an opportunity to evaluate a graviton mass analyzing
a time delay in electromagnetic waves such as supernova or gamma-ray
burst \cite{Will_14}, moreover earlier he demonstrated a possibility
to constrain a graviton mass from from gravitational wave signal
alone \cite{Will_98}.

Pulsar timing may be used to evaluate a graviton mass \cite{Lee_10}.
In the paper it was conclude that, with 90\% probability, massless
gravitons can be distinguished from gravitons heavier than $3 \times
10^{-22}$ eV (Compton wavelength $\lambda_g = 4.1 \times 10^{12}$
km), if bi-weekly observation of 60 pulsars is performed for 5 years
with a pulsar rms timing accuracy of 100 ns and if 10 year
observation of 300 pulsars with 100 ns timing accuracy would probe
graviton masses down to $3 \times 10^{-23}$ eV ($\lambda_g = 4.1
\times 10^{13}$ km). These conclusions are based on an analysis of
cross-correlation functions of gravitational wave background. An
idea to use pulsar timing for gravitational wave detection has been
proposed many years ago \cite{Sazhin_78}. An analysis of the
cross-correlation function between pulsar timing residuals of pulsar
pairs could give an opportunity to detect gravitational waves
\cite{Jenet_05,Lee_08}. If a graviton has a mass it gives an impact
on cross-correlation functions \cite{Lee_10}. However, as a first
step people have to discover stochastic GW signal and only after a
detailed analysis of cross-correlation it could help to put
constraints on a graviton mass.

Here we show that our previous results concerning the constraints on
parameters of Yukawa gravity, presented in the paper
\cite{Dusko_JCAP_13}, can be extended in the way that one could also
obtain a graviton mass bounds from the observations of trajectories
of bright stars near the Galactic Center. As it is shown below our
estimate of a graviton mass is slightly greater than the estimate
obtained  by the LIGO collaboration with the first detection of
gravitational waves from the binary black hole system. However, we
would like to note that a) our estimate is consistent with the LIGO
one; b) in principle, with analysis of trajectories of bright stars
near the Galactic Center one can obtain such a graviton mass
estimate before the LIGO report \cite{Abbott_16} about the discovery
of gravitational waves and their estimate of  a graviton mass; c) in
the future our estimate may be improved with forthcoming
observational facilities.

\section{Graviton mass estimates from S2 star orbit}

Two groups of observers are monitoring bright stars (including S2
one) to evaluate gravitational potential at the Galactic Center
\cite{Ghez_00,Rubilar_01,Schodel_02,Ghez_03,Ghez_04,Ghez_08,Gillessen_09a,
Gillessen_12,Meyer_12}.
Recently, the astrometric observations of S2 star
\cite{Gillessen_09b} were used to evaluate parameters of black hole
and to test and constrain several models of modified gravity at mpc
scales
\cite{Zakharov_07,Zakharov_09,Dusko_PRD_12,Zakharov_15,Zakharov_16}.
The simulations of the S2 star orbit around the supermassive black
hole at the Galactic Centre (GC) in Yukawa gravity
\cite{Dusko_JCAP_13} and their comparisons with the NTT/VLT
astrometric observations of S2 star \cite{Gillessen_09b} resulted
with the constraints on the range of Yukawa interaction $\lambda$,
which showed that $\lambda$ is most likely on the order of several
thousand astronomical units. However, it was not possible to obtain
the reliable constrains on the universal constant $\delta$ because
its values $0 <\delta < 1$ were highly correlated to $\lambda$,
while the values $\delta > 2$ corresponded to a practically fixed
$\lambda\sim 5000 - 6000$ AU. Such behavior of $\lambda$ indicate
that it can be used to constrain the lower bound for Compton
wavelength $\lambda_g$ of the graviton, and thus the upper bound for
its mass $m_g=h\,c/\lambda_g$, assuming Yukawa gravitational
potential of a form $\propto r^{-1}\exp(-r/\lambda_g)$ \citep[see
e.g.][]{Will_98}. The goal of this paper is to find these
constraints using chi-square test of goodness of the S2 star orbit
fits by Yukawa gravity potential (\ref{eq_yukawa}).

For that purpose and in order to obtain the reliable statistical
criterion, we had to modify the $\chi^2$ measure of goodness of the
fit given in \cite{Dusko_JCAP_13} to the following expression:
\begin{equation}
\chi^{2} =\sum\limits_{i=1}^n\left[\dfrac{\left(x_i^o -
x_i^c\right)^2}{\sigma_{xi}^2+\sigma_{int}^2}+\dfrac{\left(y_i^o -
y_i^c\right)^2}{\sigma_{yi}^2+\sigma_{int}^2}\right],
\label{eq_chi2}
\end{equation}
where $(x_i^o, y_i^o)$ is the $i$-th observed position, $(x_i^c,
y_i^c)$ is the corresponding calculated position, $n$ is the number
of observations, $\sigma_{xi}$ and $\sigma_{yi}$ are uncertainties
of observed positions, while $\sigma_{int}$ accounts for the
''intrisic dispersion'' of the data. $\sigma_{int}$ is usually
introduced whenever the observed uncertainties are not mutually
consistent, like it is the case with SN Ia data in cosmology (see
e.g. \cite{ast06,zakh12} and references therein). In our case
introduction of $\sigma_{int}$ was necessary because the astrometric
accuracy of the S2 star observations changed for more than order of
magnitude during the observational period, improving from around 10
mas at the beginning, up to around 0.3 mas at the end. As it will be
shown below, $\sigma_{int}$ will not affect the best fit values for
$\lambda$, but it will scale $\chi^2$ so that it could be used as a
proper statistics for hypothesis testing.

\begin{figure}[ht!]
\centering
\includegraphics[width=0.75\textwidth]{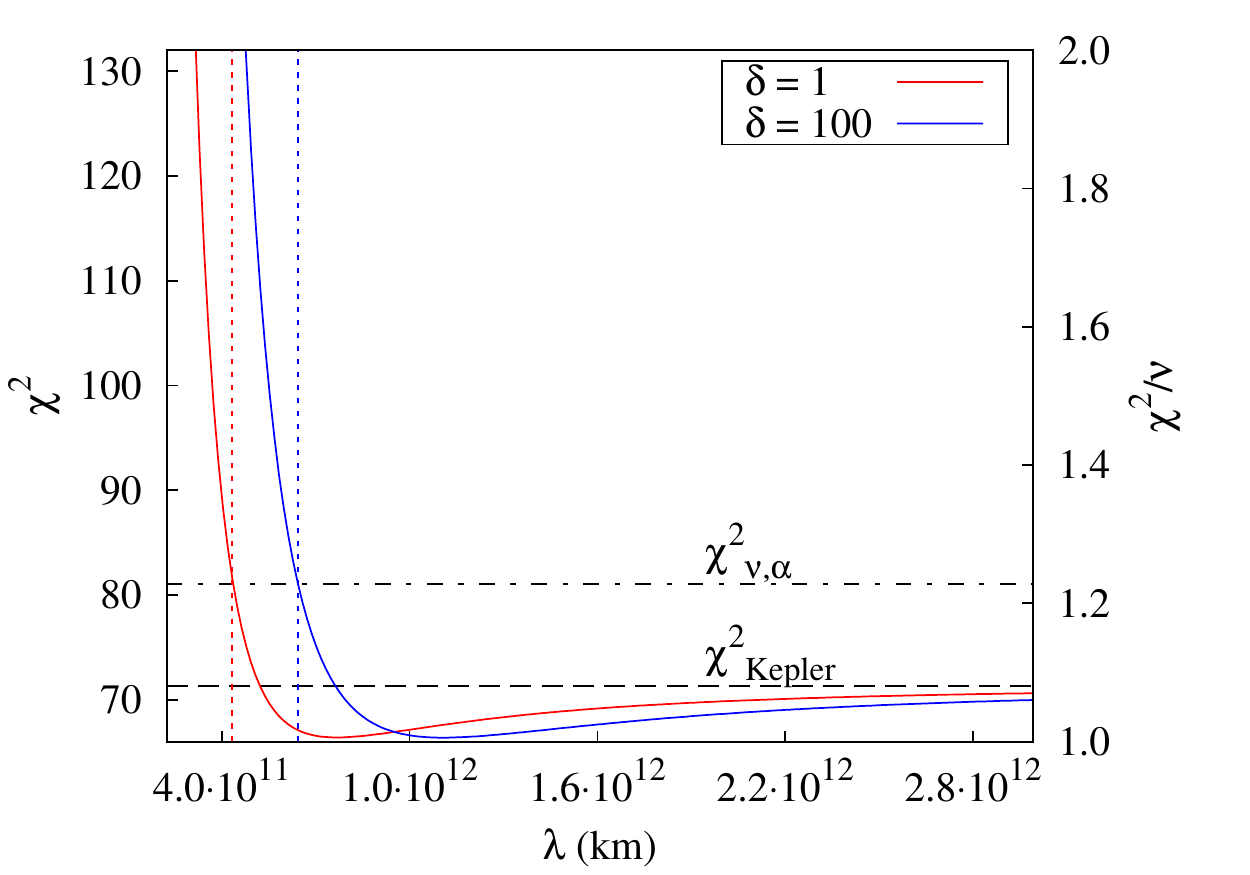}
\caption{$\chi^2$ (solid lines) as a function of Yukawa range of
interaction $\lambda$, i.e. the graviton Compton wavelength
$\lambda_g$, obtained from the fits of NTT/VLT observations of S2
star \cite{Gillessen_09b} using the gravity potential
(\ref{eq_yukawa}) for $\delta=1$ (red) and $\delta=100$ (blue). For
comparison, the value of the Keplerian fit
$\chi^2_\mathrm{Kepler}=71.34$ is also denoted by the horizontal
dashed line. The critical value for $\nu=66$ degrees of freedom and
$\alpha=0.1$ significance level, $\chi^2_{\nu,\alpha}=81.08$, is
presented by the horizontal dash-dotted line, and the upper bounds
$\lambda_x$ of the corresponding exclusion regions for $\lambda_g$
by the vertical dotted lines. The values $\lambda_g<\lambda_x$ can
be excluded with 90\% probability.} \label{fig1}
\end{figure}

We then performed the new fits of $n=70$ positions of S2 star
observed by NTT/VLT \cite{Gillessen_09b} with its simulated orbits in
Yukawa  gravity potential (\ref{eq_yukawa}), by varying $\lambda$
between 1500 and 20000 AU and assuming two values of $\delta$:
$\delta=1$ (belonging to the region where $\delta$ and $\lambda$ are
correlated) and $\delta=100$ (belonging to the region where there is
no correlation between $\delta$ and $\lambda$). In total 4
parameters were fitted: two components of initial position and two
components of initial velocity in orbital plane, and thus the number
of degrees of freedom is $\nu=66$ \cite{Dusko_JCAP_13}. Fitting is
done by minimization of $\chi^2$ given by (\ref{eq_chi2}), where
$\sigma_{int}$ is estimated from the requirement that the global
minimum of reduced $\chi^2$ over the whole range of $\lambda$ is
$\chi^2/\nu = 1$ (see Fig. \ref{fig1}). It is found that
$\sigma_{int}=1.13$ mas satisfies this requirement. The resulting
values of $\chi^2$ for $\delta=1$ and $\delta=100$ as functions of
$\lambda$ are presented in Fig. \ref{fig1} by red and blue solid
curves, respectively. As it can be seen from this figure, $\chi^2$
asymptotically approaches to the corresponding value of the
Keplerian fit $\chi^2_\mathrm{Kepler}=71.34$ (horizontal dashed line
in Fig. \ref{fig1}), for the large values of $\lambda$. Besides,
$\chi^2$ has global minimum at
$\lambda=5100\pm50\,\mathrm{AU}\approx 7.6\times
10^{11}\,\mathrm{km}$ in the case of $\delta=1$, and at
$\lambda=7400\pm50\,\mathrm{AU}\approx 1.1\times
10^{12}\,\mathrm{km}$ in the case of $\delta=100$. Thus, the
obtained results for $\lambda$ are consistent with those from
\cite{Dusko_JCAP_13}.

In the next step, we use the obtained values of $\chi^2$ to test the
null hypothesis that $\lambda$ should be at least on the order of
$10^3$ AU. For testing this hypothesis we assumed the significance
level $\alpha=0.1$ and calculated the critical value
$\chi^2_{\nu,\alpha}=81.08$ for $\nu=66$ degrees of freedom
(horizontal dash-dotted line in Fig. \ref{fig1}). As it can be seen
from Fig. \ref{fig1}, for both $\chi^2$ curves there is an exclusion
range of $\lambda$ with some upper bound $\lambda_x$ where $\chi^2 >
\chi^2_{\nu,\alpha}$, so the cases $\lambda <\lambda_x$ can be
excluded with very high probability of $1-\alpha=90\%$. In the case
of $\delta=1$ this upper exclusion bound is
$\lambda_x=2900\pm50\,\mathrm{AU}\approx 4.3\times
10^{11}\,\mathrm{km}$, while in the case of $\delta=100$ it is
$\lambda_x=4300\pm50\,\mathrm{AU}\approx 6.4\times
10^{11}\,\mathrm{km}$. Since the null hypothesis can be considered
as valid for $\lambda
>\lambda_x$, this value can be taken as the lower allowed bound for the
graviton Compton wavelength $\lambda_g > \lambda_x$. By comparing
this result with those presented in Fig. 8 from \cite{ligo16}, one
can see that it is in agreement with Solar System and LIGO
constraints on $\lambda_g$.

The corresponding upper bound for graviton mass, $m_g
=h\,c/\lambda_x$, according to the fits of the  astrometric
observations of S2 star by its simulated orbits in Yukawa gravity,
is $m_g=2.9\times 10^{-21}\ \mathrm{eV}$ in the case of $\delta=1$
and $m_g=1.9\times 10^{-21}\ \mathrm{eV}$ in the case of
$\delta=100$. These constraints are consistent with those obtained
from a gravitation wave signal GW150914 recently detected by LIGO
\cite{ligo16}, and significantly exceeds $1.2\times 10^{-22}$ eV
which represents 90\% probability limit for distinguishing massless
gravitons (predicted by General relativity) from massive gravitons
(predicted by modified gravity theories with Yukawa type of
gravitational potential) using pulsar timing experiments \cite[see
e.g.][]{Lee_10}.

\section{Conclusions}
In this paper we consider phenomenological consequences of massive
gravity and show that an analysis of bright star trajectories could
bound the graviton mass. Using simulations of the S2 star orbit
around the supermassive black hole at the Galactic Center in Yukawa
gravity \cite{Dusko_JCAP_13} and their comparisons with the NTT/VLT
astrometric observations of S2 star \cite{Gillessen_09b} we get the
constraints on the range of Yukawa interaction which showed that
$\lambda > 4.3 \times 10^{11}$ km. Taking this value as the lower
bound for the graviton Compton wavelength, we found that the
corresponding most likely upper bound for graviton mass is $m_g <
2.9 \times 10^{-21}\ \mathrm{eV}$. This result is consistent with
the constraints obtained from a gravitation wave signal GW150914
recently detected by LIGO, and significantly exceeds 90\%
probability limit for distinguishing massless gravitons from massive
gravitons using pulsar timing experiments.

{ Planned observations of trajectories of bright stars near the
Galactic Center with GRAVITY \cite{Blind_15}, E-ELT \cite{EELT} and
TMT \cite{TMT} may improve the discussed estimates of graviton
masses}.

\acknowledgments

A. F. Z. thanks  a senior scientist fellowship of Chinese Academy of
Sciences for a partial support and prof. K. Lee (KIAA PKU) for
useful discussion of an opportunity to constrain a graviton mass
with observational data from pulsar timing. A. F. Z. thanks also NSF
(HRD-0833184) and NASA (NNX09AV07A) at NASA CADRE and NSF CREST
Centers (NCCU, Durham, NC, USA) for a partial support. P.J., D.B.
and V.B.J. wish to acknowledge the support by the Ministry of
Education, Science and Technological Development of the Republic of
Serbia through the project 176003 ''Gravitation and the large scale
structure of the Universe''.


\begin{thebibliography}{99}

\bibitem[\protect\citeauthoryear{Abbott et al.}{2016}]{Abbott_16}
B.~P.~Abbott et al., (LIGO Scientific Collaboration and Virgo
Collaboration), \emph{Observation of Gravitational Waves from a
Binary Black Hole Merger}, Phys. Rev. Lett., {\bf 116} (2016)
061102.

\bibitem[\protect\citeauthoryear{Abbott et al.}{2016}]{ligo16}
B.~P.~Abbott et al., (LIGO Scientific Collaboration and Virgo
Collaboration), \emph{Tests of general relativity with GW150914},
LIGO Document P1500213-v27, arXiv:1602.03841 (2016).


\bibitem[\protect\citeauthoryear{Fierz and Pauli}{1939}]{FP} M.~Fierz and
W.~Pauli,
\emph{On Relativistic Wave Equations for Particles of Arbitrary Spin
in an Electromagnetic Field}, Proc. of the Royal Society of London
{\bf A173} (1939) 211.

\bibitem[\protect\citeauthoryear{Will}{1998}]{Will_98}
C.~Will, \emph{Bounding the mass of the graviton using
gravitational-wave observations of inspiralling compact binaries},
Phys.  Rev. D  {\bf 57} (1998) 2061; [gr-qc/9709011].


\bibitem[\protect\citeauthoryear{Will}{2014}]{Will_14}
C.~Will, \emph{The Confrontation between General Relativity and
Experiment}, Living Reviews in Relativity,  {\bf 17} (2014) 4 ;
arXiv:1403.7377v1[gr-qc].


\bibitem[\protect\citeauthoryear{Anderson et al.}{1998}]{Anderson_98}
J.~D.~Anderson, P.~A.~Laing, E.~L.~Lau, A.~S.~Liu, M.~M.~Nieto and
S.~G.~Turyshev,  \emph{Indication, from Pioneer 10/11, Galileo, and
Ulysses Data, of an Apparent Anomalous, Weak, Long-Range
Acceleration}, Phys. Rev. Lett., {\bf 81} (1998) 2858.

\bibitem[\protect\citeauthoryear{Anderson et al.}{2008}]{Anderson_08}
J.~D.~Anderson, J.~K.~Campbell, J.~E.~Ekelund, J.~Ellis and
J.~F.~Jordan,  \emph{Anomalous Orbital-Energy Changes Observed
during Spacecraft Flybys of Earth}, Phys. Rev. Lett., {\bf 100}
(2008) 091102.

\bibitem[\protect\citeauthoryear{Turyshev et al.}{2012}]{Turyshev_12}
S.~G.~Turyshev, V.~T.~Toth, G.~Kinsella, S.~C.~Lee, S.~M.~Lok and
J.~Ellis, \emph{Support for the Thermal Origin of the Pioneer
Anomaly}, Phys. Rev. Lett. {\bf 108} (2012) 241101.

\bibitem[\protect\citeauthoryear{Zakharov}{1970}]{Zakharov_70}
V.~I.~Zakharov, \emph{Linearized Gravitation Theory and the Graviton
Mass}, JETP Letters {\bf 12} (1970) 447.

\bibitem[\protect\citeauthoryear{van Dam \& Veltman}{1970}]{van_Dam_70}
 H.~van Dam and M.~Veltman,  \emph{Massive and mass-less Yang-Mills and gravitational fields}, Nucl. Phys. B {\bf 22}  (1970) 397.

\bibitem[\protect\citeauthoryear{Iwasaki}{1970}]{Iwasaki_70}
Y.~Iwasaki, \emph{Consistency Condition for Propagators}, Phys. Rev.
D {\bf 2}   (1970) 2255.



\bibitem[\protect\citeauthoryear{Rubakov and Tinyakov}{2008}]{Rubakov_08}
V.~A.~Rubakov and P.~G.~Tinyakov, \emph{Infrared-modified gravities
and massive gravitons}, Physics -- Uspekhi {\bf 51} (2008) 759.

\bibitem[\protect\citeauthoryear{Vainshtein}{1972}]{Vainshtein_72}
A.~I.~Vainshtein, \emph{To the problem of nonvanishing gravitation
mass}, Phys. Lett. B {\bf 39}  (1972) 393.

\bibitem[\protect\citeauthoryear{Visser}{1998}]{Visser_98}
M.~Visser,  \emph{Mass for the Graviton}, Gen. Rel. Grav.  {\bf 30},
1717 (1998).

\bibitem[\protect\citeauthoryear{Dvali et al.}{2000}]{Dvali_00}
G.~Dvali, G.~Gabadadze and M.~Porrati, \emph{Metastable gravitons
and infinite volume extra dimensions}, Phys. Lett. B {\bf 485}
 (2000) 112.

\bibitem[\protect\citeauthoryear{Kogan et al.}{2001}]{Kogan_01}
I.~I.~Kogan, S.~Mouslopoulos, A.~Papazoglou and G.~G.~Ross,
\emph{Multigravity in six dimensions: Generating bounces with flat
positive tension branes}, Nucl. Phys. B {\bf 595} (2001)  225.

\bibitem[\protect\citeauthoryear{Deffayet et al.}{2002}]{Deffayet_02}
C.~Deffayet, G.~Dvali, G.~Gabadadze and A.~Vainshtein,
\emph{Nonperturbative continuity in graviton mass versus
perturbative discontinuity}, Phys. Rev. D {\bf 65}, 044026 (2002).

\bibitem[\protect\citeauthoryear{Finn \& Sutton}{2002}]{Finn_02}
L.~S.~Finn and P.~J.~Sutton, \emph{Bounding the mass of the graviton
using binary pulsar observations}, Phys. Rev. D {\bf 65} (2002)
044022.

\bibitem[\protect\citeauthoryear{Damour, Kogan \& Papazoglou}{2003}]{Damour_03}
T.~Damour, I.~I.~Kogan and A.~Papazoglou, \emph{Spherically
symmetric spacetimes in massive gravity}, Phys. Rev. D {\bf 67},
064009 (2003).


\bibitem[\protect\citeauthoryear{Maggiorre}{2008}]{Maggiorre_08}
M.~Maggiorre,  \emph{Gravitational Waves}, Vol. VI, Oxford, Oxford
University Press (2008).

\bibitem[\protect\citeauthoryear{Hinterbichler}{2012}]{Hinterbichler_12}
K.~Hinterbichler, \emph{Theoretical aspects of massive gravity},
Rev. Mod. Phys. {\bf 84} (2012) 671.

\bibitem{capo09a}
S.~Capozziello, A.~Stabile and A.~Troisi, \emph{A general solution
in the Newtonian limit of f(R) - gravity}, {Mod. Phys. Lett.} A {\bf
24}  (2009) 659.

\bibitem[\protect\citeauthoryear{Goldhaber and Nieto}{1974}]{Nieto_74}
A.~S.~Goldhaber and M.~M.~Nieto, \emph{Mass of the graviton}, Phys.
Rev. D {\bf 9}  (1974) 1119.

\bibitem[\protect\citeauthoryear{Freund et al.}{1969}]{Freund_69}
P.~G.~O.~Freund, A.~Maneshwari and E.~Schonberg, \emph{Finite-Range
Gravitation}, Astrophys. J. {\bf 157}  (1969) 857.

\bibitem[\protect\citeauthoryear{Logunov \& Mestvirishvili}{1984}]{Logunov_84}
A.~A.~Logunov and M.~A.~Mestvirishvili, \emph{Relativistic theory of
gravitation}, Theor. Math. Phys. {\bf 61} (1984) 1170.


\bibitem[\protect\citeauthoryear{Gershtein, Logunov \& Mestvirishvili}{2006}]{Gershtein_06}
S.~S.~Gershtein, A.~A.~Logunov and M.~A.~Mestvirishvili,
\emph{Gravitational field self-limitation and its role in the
Universe}, Physics -- Uspekhi, {\bf 49} (2006) 1179.

\bibitem[\protect\citeauthoryear{Gershtein, Logunov \& Mestvirishvili}{2008}]{Gershtein_08}
S.~S.~Gershtein, A.~A.~Logunov and M.~A.~Mestvirishvili,
\emph{General relativity and the Schwarzschild singularity}, Phys.
Part. Nucl.{\bf 39} (2008)  1.

\bibitem[\protect\citeauthoryear{Goldhaber and Nieto}{2010}]{Nieto_10}
A.~S.~Goldhaber and M.~M.~Nieto, \emph{Photon and graviton mass
limits}, Rev. Mod. Phys. {\bf 82}  (2010) 939.

\bibitem[\protect\citeauthoryear{Kopeikin}{2001}]{Kopeikin_01}
S.~M.~Kopeikin, \emph{Testing the relativistic effect of the
propagation of gravity by very long baseline interferometry},
Astrophys. J. Lett. {\bf 556}  (2001) L1; [gr-qc/0105060].

\bibitem[\protect\citeauthoryear{Kopeikin}{2003}]{Kopeikin_03}
E.~B.~Fomalont and S.~M.~Kopeikin, \emph{The measurement of the
light deflection from Jupiter: experimental results}, Astrophys. J.
{\bf 598} (2003) 704; [astro-ph/0302294].

\bibitem[\protect\citeauthoryear{Kopeikin}{2004}]{Kopeikin_04}
S.~M.~Kopeikin, \emph{The speed of gravity in general relativity and
theoretical interpretation of the Jovian deflection experiment},
Class. Quant. Grav. {\bf 21}, 3251 (2004), [gr-qc/0310059].

\bibitem[\protect\citeauthoryear{Kopeikin}{2004}]{Kopeikin_05}
S.~M.~Kopeikin, \emph{Comment on 'Model-dependence of Shapiro time
delay and the "speed of gravity/speed of light" controversy'},
Class. Quantum Grav. {\bf 22}  (2005) 5181; [gr-qc/0501048].


\bibitem[\protect\citeauthoryear{Kopeikin}{2004}]{Kopeikin_06}
S.~M.~Kopeikin, \emph{Comments on "On the Speed of Gravity and the
Jupiter/quasar Measurement" by S. Samuel}, Int. J. Mod. Phys. D {\bf
15}  (2006) 273; [gr-qc/0501001].

\bibitem[\protect\citeauthoryear{Talmadge et al.}{1988}]{Talmadge_88}
C.~Talmadge, J.-P.~Berthias, R.~W.~Hellings and E.~M.~Standish,
\emph{Model-independent constraints on possible modifications of
Newtonian gravity}, Phys. Rev. Lett. {\bf 61}  (1988) 1159.

\bibitem{Dusko_JCAP_13} D.~Borka, P.~Jovanovi{\'c}, V.~Borka Jovanovi{\'c},
A.~F.~Zakharov, \emph{Constraining the range of Yukawa gravity
interaction from S2 star orbits}, JCAP {\bf 11} (2013) 050.

\bibitem[\protect\citeauthoryear{Lee et al.}{2010}]{Lee_10}
K.~J.~ Lee, F.~A.~Jenet, R.~H.~Price, N.~Wex and M.~Kramer,
\emph{Detecting Massive Gravitons Using Pulsar Timing Arrays},
Astrophys. J. {\bf 722}  (2010) 1589.

\bibitem[\protect\citeauthoryear{Sazhin}{1978}]{Sazhin_78}
M.~V.~Sazhin, \emph{Opportunities for detecting ultralong
gravitational waves}, Sov. Astron. {\bf 22}  (1978) 36.

\bibitem[\protect\citeauthoryear{Jenet et al.}{2005}]{Jenet_05}
F.~A.~Jenet, G.~B.~Hobbs, K.~J.~Lee and R.~N.~Manchester,
\emph{Detecting the Stochastic Gravitational Wave Background Using
Pulsar Timing}, Astrophys. J. {\bf 625}  (2005) L123.

\bibitem[\protect\citeauthoryear{Lee et al.}{2008}]{Lee_08}
K.~J.~Lee, F.~A.~Jenet and R.~H.~Price,  \emph{Pulsar Timing as a
Probe of Non-Einsteinian Polarizations of Gravitational Waves},
Astrophys. J. {\bf 685}  (2008) 1304.

\bibitem[\protect\citeauthoryear{Ghez et al.}{2000}]{Ghez_00}
A.~M.~Ghez, M.~Morris, E.~E.~Becklin, A.~Tanner and
T.~Kremenek,\emph{The accelerations of stars orbiting the Milky
Way's central black hole}, Nature {\bf 407} (2000) 349.

\bibitem[\protect\citeauthoryear{Rubilar and Eckart}{2001}]{Rubilar_01}
G.~F.~Rubilar and A.~Eckart, \emph{Periastron shifts of stellar
orbits near the Galactic Center}, Astron. Astrophys. {\bf 374},
 (2001) 95.

\bibitem[\protect\citeauthoryear{Sch\"odel et al.}{2002}]{Schodel_02}
 R.~Sch\"odel, T.~Ott, R.~Genzel  {et al.}, \emph{A star in a 15.2-year orbit around the supermassive black hole at the centre of the Milky Way},
  Nature {\bf 419}, 694
 (2002).

\bibitem[\protect\citeauthoryear{Ghez et al.}{2003}]{Ghez_03}
A.~M.~Ghez, G. Duch\^{e}ne, K. Matthews  {et al.}, \emph{The First
Measurement of Spectral Lines in a Short-Period Star Bound to the
Galaxy's Central Black Hole: A Paradox of Youth}, Astrophys. J.
Lett. {\bf 586} (2003) L127.

\bibitem[\protect\citeauthoryear{Ghez et al.}{2004}]{Ghez_04}
A.~M.~Ghez, S.~ A.~Wright,  K. Matthews {et al.}, \emph{Variable
Infrared Emission from the Supermassive Black Hole at the Center of
the Milky Way}, Astrophys. J. Lett. {\bf 601} (2004) L159.

\bibitem[\protect\citeauthoryear{Ghez et al.}{2008}]{Ghez_08}
A.~M.~Ghez, S.~Salim, N.~N.~Weinberg {et al.}, \emph{Measuring
Distance and Properties of the Milky Way's Central Supermassive
Black Hole with Stellar Orbits}, Astrophys. J. {\bf 689}  (2008)
1044.

\bibitem[\protect\citeauthoryear{Gillessen et al.}{2009}]{Gillessen_09a}
S.~Gillessen, F.~Eisenhauer, S.~Trippe {et al.}, \emph{Monitoring
Stellar Orbits Around the Massive Black Hole in the Galactic
Center}, Astrophys. J. {\bf 692}  (2009) 1075.

\bibitem[\protect\citeauthoryear{Gillessen et al.}{2012}]{Gillessen_12}
S.~Gillessen, R.~Genzel, T.~K.~Fritz {et al.}, \emph{A gas cloud on
its way towards the supermassive black hole at the Galactic Centre},
Nature, {\bf 481}  (2012) 51.

\bibitem[\protect\citeauthoryear{Meyer et al.}{2012}]{Meyer_12}
 L.~Meyer, A.~M.~Ghez, R.~Sch\"odel {et al.}, \emph{The Shortest-Known-Period Star Orbiting Our Galaxy's Supermassive Black Hole}, Science, {\bf 338}  (2012) 84.

\bibitem[\protect\citeauthoryear{Gillessen et al.}{2009}]{Gillessen_09b}
S.~Gillessen, F.~Eisenhauer, T.~K.~Fritz {et al.}, \emph{The Orbit
of the Star S2 Around SGR A* from Very Large Telescope and Keck
Data}, Astrophys. J. {\bf 707}  (2009) L114.


\bibitem{Zakharov_07}
A.~F.~Zakharov, A.~A.~Nucita, F.~De Paolis and G.~Ingrosso,
\emph{Apoastron Shift Constraints on Dark Matter Distribution at the
Galactic Center}, Phys. Rev. D {\bf 76}  (2007) 062001.

\bibitem{Zakharov_09}
A.~F.~Zakharov, S.~Capozziello, F.~De Paolis, G.~Ingrosso and
A.~A.~Nucita, \emph{ The Role of Dark Matter and Dark Energy in
Cosmological Models: Theoretical Overview}, {Space Sci. Rev.} {\bf
48}  (2009) 301.

\bibitem{Dusko_PRD_12}
D.~Borka,  P.~Jovanovi\'c,  V.~Borka Jovanovi\'c  and
A.~F.~Zakharov, \emph{Constraints on $R^n$ gravity from precession
of orbits of S2-like stars}, {Phys. Rev. D}  {\bf 85} (2012) 124004.

\bibitem{Zakharov_15}
A.~F.~Zakharov, \emph{Possible Alternatives to the Supermassive
Black Hole at the Galactic Center}, J. Astrophys. Astron., {\bf 36}
(2015) 539.

\bibitem{Zakharov_16}
A.~F.~Zakharov, \emph{Is there an ordinary supermassive black hole
at the Galactic Center?}, in Gravitation, Astrophysics, and
Cosmology - Proc. Twelfth Asia-Pacific International Conference,
edited by Jong-Ping Hsu et al., World Scientific Publ., Singapore,
(2016) p. 176.

\bibitem{ast06} P.~Astier, J.~Guy,  N.~Regnault et al., \emph{The Supernova Legacy Survey: measurement of $\Omega_M$, $\Omega_\Lambda$
and $w$ from the first year data set}, Astron. Astrophys. {\bf 447}
(2006) 31.

\bibitem{zakh12} A.~F.~Zakharov and V.~N.~Pervushin, \emph{Conformal cosmological model and SNe Ia data
}, Phys. Atom. Nucl. {\bf 75} (2012) 1418.

\bibitem[\protect\citeauthoryear{Blind et al.}{2015}]{Blind_15}
N.~Blind, F. Eisenhauer, S. Gillessen  {et al.}, \emph{GRAVITY: the
VLTI 4-beam combiner for narrow-angle astrometry and interferometric
imaging}, arXiv:1503.07303 [astro-ph.IM].

\bibitem[\protect\citeauthoryear{Ardeberg et al.}{2014}]{EELT}
A.~Ardeberg, J.~Bergeron, A.~Cimatti {et al.}, \emph{An Expanded
View of the Universe Science with the European Extremely Large
Telescope}, ed. by  M. Lyubenova and M. Kissler-Patig, ESO, Munich.

\bibitem[\protect\citeauthoryear{Skidmore et al.}{2014}]{TMT}
W.~Skidmore {et al.}, \emph{Thirty Meter Telescope Detailed Science
Case: 2015 (TMT Observatory Corporation)}.

\end{thebibliography}
\end{document}